\documentclass[twocolumn,english,prc,nofootinbib,floatfix,showpacs]{revtex4}
\usepackage[T1]{fontenc}
\usepackage[latin9]{inputenc}
\usepackage{amsmath}
\usepackage{graphicx}
\usepackage{amssymb}
\usepackage{esint}

\makeatletter

\newcommand{\noun}[1]{\textsc{#1}}

\@ifundefined{textcolor}{}
{%
 \definecolor{BLACK}{gray}{0}
 \definecolor{WHITE}{gray}{1}
 \definecolor{RED}{rgb}{1,0,0}
 \definecolor{GREEN}{rgb}{0,1,0}
 \definecolor{BLUE}{rgb}{0,0,1}
 \definecolor{CYAN}{cmyk}{1,0,0,0}
 \definecolor{MAGENTA}{cmyk}{0,1,0,0}
 \definecolor{YELLOW}{cmyk}{0,0,1,0}
 }

\usepackage{graphics}\usepackage{epsfig}\usepackage{amsfonts}
\usepackage{bm}
\@ifundefined{definecolor}
 {\usepackage{color}}{}
\newcommand{\pr}[1]{{\sc{\lowercase{#1}}}}

\makeatother

\usepackage{babel}

\begin{document}

\title{Collective vibrational states with fast iterative QRPA method}

\author{B.~G.~Carlsson$^{1}$, J.~Toivanen$^{2}$, A. Pastore$^{3}$ }

\affiliation{$^{1}$ Division of Mathematical Physics, LTH, Lund University, Post
Office Box 118, S-22100 Lund, Sweden}

\email{gillis.carlsson@matfys.lth.se}

\affiliation{$^{2}$ Department of Physics, University of Jyväskylä, P.O. Box
35 (YFL) FI-40014, Finland}

\affiliation{$^{3}$Universit\'e de Lyon, Universit\'e Lyon 1, CNRS/IN2P3 Institut
de Physique Nucl\'eaire de Lyon, F-69622 Villeurbanne cedex, France}

\date{\today}
\begin{abstract}
An iterative method we previously proposed to compute nuclear strength
functions \cite{TOI10} is developed to allow it to accurately calculate
properties of individual nuclear states. The approach is based on
the quasi-particle-random-phase approximation (QRPA) and uses an iterative
non-hermitian Arnoldi diagonalization method where the QRPA matrix
does not have to be explicitly calculated and stored. The method gives
substantial advantages over conventional QRPA calculations with regards
to the computational cost. The method is used to calculate excitation
energies and decay rates of the lowest lying $2^{+}$ and $3^{-}$
states in Pb, Sn, Ni and Ca isotopes using three different Skyrme
interactions and a separable gaussian pairing force.
\end{abstract}


\maketitle

\section{Introduction}

The goal of nuclear structure theory is to be able to predict and
model the physics of the atomic nucleus. This involves the ground-state
properties, as well as different modes of excitation and decay. One
of the possible methods to compute excited states in nuclei is based
on the quasi-particle-random-phase approximation (QRPA) \cite{Ring80}.
This approach can be derived by considering the linear response of
a nucleus when perturbed by an external field. From the response one
can extract information about excited nuclear states and cross sections
for nuclear reactions. The QRPA approach is particularly interesting
in connection with nuclear density-functional theory (DFT) as the
method can be applied also when starting from a density functional.
In order for the QRPA method to be practical, it is very important
to implement it in ways that have low computational costs. For phenomenological
DFT approaches, a low computational cost would allow dynamical properties
to be considered when fine tuning model parameters. A numerically
efficient method is also essential for applications to deformed and
heavy nuclei which are otherwise prohibited by the time and memory
required to construct and diagonalize the large QRPA matrix. 

Two recent solution methods address these issues. The Finite Amplitude
Method (FAM) \cite{FAM_Nakatsukasa,FAM_Avogadro} generates the response
of a nucleus to an external field by solving the linear response equations
iteratively for each requested external field frequency.\emph{ }FAM
furthermore uses the same mean fields as in the Hartree-Fock-Bogoliubov
(HFB) ground-state calculation and employs finite differences to linearize
the equations of motion. In its current form FAM uses a smoothing
method to improve stability and therefore one cannot easily extract
the exact QRPA eigenamplitudes. The same is true for the iterative
Arnoldi method \cite{TOI10} which is able to provide smoothened QRPA
strength functions and their energy weighed moments, but does not
generate accurate individual states. A common aspect of both methods
is however their ability to generate partial solutions of the full
QRPA problem with reduced computational effort. 

The purpose of this paper is to generalize the Arnoldi method which
we previously developed for iterative calculations of RPA strength
functions. The generalization involves modifying the method so that
it becomes possible to not only compute strength functions but also
sets of individual excited states with high accuracy. As a first step
the new method is applied to the calculation of excitation energies
and decay rates of the lowest lying $2^{+}$ and $3^{-}$ states in
several isotope and one isotone chain. Particular focus is given to
the region around double-magic $^{208}$Pb where new experiments are
currently planned \cite{experiments}. 

This paper is organized as follows: in Sec. II the QRPA formalism
is briefly reviewed and specific aspects of our formulation are discussed.
In Sec III the computational cost and accuracy of the method is evaluated.
In Sec. IV the method is applied to the calculation of energies and
transition probabilities of the lowest $J^{\pi}=2^{+}$ and $J^{\pi}=3^{-}$
states in a selection of semi-magic even-even nuclei. Finally conclusions
are given in section V.

\section{QRPA in terms of fields }

The iterative method is based on the QRPA equations \cite{RPA_from_THF,RPA-densdep,B-R,Ring80}
which can be derived by starting from time-dependent HFB theory. Here
we present the main parts of the derivation, highlighting aspects
relevant to the iterative formulation. In cases where the expressions
are not fully defined we use notation consistent with Ref.~\cite{Ring80}. 

The QRPA equations can be derived by considering a general time-dependent
wavefunction which is oscillating between the ground state and an
excited state with excitation energy $\hbar\omega$

\[
\left|\psi\left(t\right)\right\rangle =e^{-itE_{gs}/\hbar}C_{gs}\left|\psi_{g.s.}\right\rangle +e^{-it\left(E_{gs}+\hbar\omega\right)/\hbar}C_{exc}\left|\psi_{exc.}\right\rangle .\]
We limit the consideration to small-amplitude oscillations around
the ground state so that the corresponding generalized density-matrix
${\cal R}$ \cite{Ring80} can be expanded to first order in $C_{exc}$

\[
{\cal R}(t)\simeq{\cal R}_{gs}+\mbox{e}^{-i\omega t}\tilde{{\cal R}}+\mbox{e}^{i\omega t}\tilde{{\cal R}}^{\dagger}.\]

In order to make use of the time-dependent-Hartree-Fock-Bogoliubov
(TDHFB) equations of motion, it is desired that the time-dependent
density should be a HFB density at all times (i.e., it should be a
projector ${\cal R}^{2}={\cal R}$). We further assume that ${\cal R}_{gs}$
can be approximated with the ground-state HFB density. Then the most
general approximation for the transition density which ensures that
${\cal R}\left(t\right)$ a projector for small-amplitude vibrations
involves both the forward $\tilde{Z}$ and the backward $\tilde{Z}'^{\dagger}$
going amplitudes 

\begin{align}
\mathcal{U}^{\dagger}\tilde{{\cal R}}\mathcal{U} & =-C_{gs}^{*}C_{exc}\left\langle \psi_{g.s.}\right|\left(\begin{array}{cc}
\alpha\alpha^{\dagger} & \alpha\alpha\\
\alpha^{\dagger}\alpha^{\dagger} & \alpha^{\dagger}\alpha\end{array}\right)\left|\psi_{exc.}\right\rangle \label{eq:trans}\\
 & \simeq\left(\begin{array}{cc}
0 & \tilde{Z}\\
\tilde{Z}'^{\dagger} & 0\end{array}\right).\nonumber \end{align}
In this expression we have made use of the matrix

\[
\mathcal{U}=\left(\begin{array}{cc}
U & V^{*}\\
V & U^{*}\end{array}\right),\]
written in terms of the $U$ and $V$ pairing matrices \cite{Ring80}
related to the HFB ground state as well as the matrices of quasiparticle
operators $\left[\alpha\alpha^{\dagger}\right]_{ij}=\alpha_{i}\alpha_{j}^{\dagger}$.
It should be noted that if the wavefunctions in the beginning were
taken as HFB vacuums one would not obtain any backward going amplitudes,
as can be seen from Eq.~(\ref{eq:trans}) by inserting the HFB ground
state. Thus it is the assumption of the transition density $\tilde{{\cal R}}$
being as general as allowed by the ${\cal R}^{2}={\cal R}$ criteria,
which allows for the existence of the implicitly defined correlated
ground state. 

Inserting the expression for ${\cal R}(t)$ into the TDHFB equations
of motion $i\hbar\frac{d{\cal R}(t)}{dt}=\left[{\cal H},{\cal R}(t)\right]$
\cite{B-R} and taking the small-amplitude limit leads to the QRPA
equation

\[
\hbar\omega\tilde{{\cal R}}\simeq[{\cal H}[{\cal R}_{gs}],\tilde{{\cal R}}]+[{\cal H}^{1}[\tilde{{\cal R}}],{\cal R}_{gs}].\]
In this expression the hermicity property ${\cal H}^{1}[\tilde{{\cal R}}]=\left({\cal H}^{1}[\tilde{{\cal R}}^{\dagger}]\right)^{\dagger}$
of the effective interaction was assumed and the time-dependent fields
${\cal H}(t)=\partial\mathcal{E}/\partial\mathcal{R}$ were expanded
around the ground-state value \[
{\cal H}[{\cal R}]\simeq{\cal H}[{\cal R}_{gs}]+{\cal H}^{1}[{\cal R}-{\cal R}_{gs}].\]
This expansion is taken to first order in the transitional fields,
which is enough for small-amplitude vibrations and leads to

\[
{\cal H}^{1}[\tilde{R}]=\left(\begin{array}{cc}
\tilde{h} & \tilde{\Delta}\\
\tilde{\Delta}'^{\dagger} & -\tilde{h}^{T}\end{array}\right),\]
where

\begin{align*}
\tilde{h}_{\mu\nu}= & \sum_{\pi\lambda}\left.\frac{\partial h_{\mu\nu}}{\partial\rho_{\pi\lambda}}\right|_{\rho_{g.s.}}\tilde{\rho}_{\pi\lambda},\\
\tilde{\Delta}_{\mu\nu}= & \frac{1}{2}\sum_{kl}v_{\mu\nu kl}^{pp}\tilde{\kappa}_{kl},\\
\tilde{\Delta}'{}_{\mu\nu}^{*}= & \frac{1}{2}\sum_{kl}v_{\mu\nu kl}^{pp*}\tilde{\kappa}_{kl}^{*}.\end{align*}
In our case with a density-independent pairing interaction it is only
the $h=\partial\mathcal{E}/\partial\rho$ field which is non-linear
in the densities and becomes linearized. With a density-dependent
pairing interaction the $\tilde{\Delta},\tilde{\Delta}'$ fields would
also have to be linearized and would give an additional contribution
to the $\tilde{h}$ field. 

Inserting the expressions for fields and densities into the QRPA equation,
and multiplying from the left with $\mathcal{U}^{\dagger}$ and from
the right with $\mathcal{U}$, gives a system of equations for the
unknown excitation energies $\hbar\omega$ and the $\tilde{Z}$ and
$\tilde{Z}'$ amplitudes:

\begin{eqnarray*}
\hbar\omega\tilde{Z} & = & E\tilde{Z}+\tilde{Z}E+\tilde{W},\\
-\hbar\omega\tilde{Z}'^{\dagger} & = & E\tilde{Z}'^{\dagger}+\tilde{Z}'^{\dagger}E+\tilde{W}'^{\dagger}.\end{eqnarray*}
In this equation, $E$ denotes a diagonal matrix of positive quasi-particle
energies and the $W$ matrices depend on the linearized fields 

\begin{eqnarray}
\tilde{W} & = & U^{\dagger}\tilde{h}V^{*}+U^{\dagger}\tilde{\Delta}U^{*}+V^{\dagger}\tilde{\Delta}'^{\dagger}V^{*}-V^{\dagger}\tilde{h}^{T}U^{*},\nonumber \\
\tilde{W}'^{\dagger} & = & V^{T}\tilde{h}U+V^{T}\tilde{\Delta}V+U^{T}\tilde{\Delta}'^{\dagger}U-U^{T}\tilde{h}^{T}V,\label{eq:W}\end{eqnarray}
which can be expressed in terms of the transition densities

\begin{eqnarray}
\tilde{\rho}=U\tilde{Z}V^{T}+V^{*}\tilde{Z}'^{\dagger}U^{\dagger},\nonumber \\
\tilde{\kappa}=U\tilde{Z}U^{T}+V^{*}\tilde{Z}'^{\dagger}V^{\dagger},\nonumber \\
\tilde{\kappa}'^{\dagger}=V\tilde{Z}V^{T}+U^{*}\tilde{Z}'^{\dagger}U^{\dagger}.\label{eq:rhos}\end{eqnarray}

It is instructive to look back and consider the approximations used
in the derivation of these equations. The main approximations appear
to be the use of the TDHFB equations of motion, which restricts us
to the consideration of excited states connected by two quasiparticle
operators to the ground state and the assumption that the ground-state
density can be approximated with the density of the HFB ground state. 

As an example to illustrate the iteration procedure, we neglect spin
and isospin and consider a term in the energy of the form \[
\mathcal{E}\left[\rho\right]=\int\rho^{\alpha+2}\left(\vec{r}\right)d\vec{r},\]
which gives the linearized field \[
\tilde{h}_{im}=\left(\alpha+2\right)\left(\alpha+1\right)\int\phi_{i}^{*}\left(\vec{r}\right)\left(\rho{}_{g.s.}^{\alpha}\left(\vec{r}\right)\tilde{\rho}\left(\vec{r}\right)\right)\phi_{m}\left(\vec{r}\right)d\vec{r}.\]
In this case, the action of the QRPA matrix on an eigenvector can
be calculated in three steps. 

The first step is to generate the densities $\tilde{\rho}$ according
to Eq.~(\ref{eq:rhos}) and expressing them in $r$ space. For the
next step, $\tilde{h}$ is calculated as above. Alternatively, in
the case of a density-independent interaction, where fields are already
linear in densities ${\cal H}^{1}[\tilde{\mathcal{R}}]={\cal H}[\tilde{\mathcal{R}}]$,
this can be achieved using the HFB mean-field routines for calculating
matrix elements. Finally multiplying the fields with $U$ and $V$
matrices as in Eq. (\ref{eq:W}) one obtains the $W$ matrices. 

The main advantage of expressing the equations in this form is that
calculating and storing two-body matrix elements can be avoided and
instead one can rely on the expressions for HFB fields. The price
to pay is that the densities and integrals for the matrix elements
of the fields are recalculated for each matrix vector product, in
the same way as when performing the HFB iterations to find the ground
state. Thus it is important to investigate how many iterations i.e.
matrix-vector products are needed in order to obtain acceptable convergence,
and whether the iteration procedure introduces numerical errors which
could lead to instabilities.

\section{accuracy and efficiency of the method}

The iterative QRPA solver is implemented by extending the program
\textsf{\pr{HOSPHE}} (v1.02) \cite{HOSPHE} and will be included
in the next published version of the program. This code uses a spherical
harmonic oscillator basis and takes advantage of the Wigner-Eckart
theorem in order to work with angular momentum reduced quantities.
The use of reduced quantities keeps the HFB and QRPA dimensions small
and makes the code a useful tool for testing different calculational
methods.

In order to verify that the QRPA implementation is correct, a comparison
is made with a recent QRPA implementation based on the \pr{HFBTHO}
code \cite{HFBTHO}. This code is able to treat axially deformed nuclei
and its QRPA implementation is based on the traditional diagonalization
of a large QRPA matrix. Therefore, applications of this code are limited
to cases where dimensions can be kept within manageable limits. 

As a test case we consider the nucleus $_{\,\,\,8}^{18}\mathrm{O}_{10}$
and compare the ground-state energy and energies of the QRPA excitations
obtained in both codes. In order to have a benchmark result that is
useful for testing future QRPA codes we list values obtained from
both codes in Table \ref{tab:QRPA-Benchmark}. %
\begin{table}[t]
\begin{tabular}{l c c} 
\hline \hline
Quantity&	\pr{HOSPHE}&	\pr{HFBTHO}+QRPA \cite{HFBTHO}\\ 
\hline 
$E_{HFB}$ & {\bf-131.6770225}\textcolor{red}{32} &{\bf-131.6770225}\textcolor{red}{19} \\
\hline
$E_{exc}\left(0^{+}\right)$ &{\bf20.49599}\textcolor{red}{056}    &{\bf20.49599}\textcolor{red}{7} \\
$E_{exc}\left(1^{-}\right)$ &{\bf14.020}\textcolor{red}{98740}    &{\bf14.020}\textcolor{red}{85} \\
$E_{exc}\left(2^{+}\right)$ &{\bf8.69120}\textcolor{red}{0427}    &{\bf8.69120}\\
$E_{exc}\left(3^{-}\right)$ &{\bf12.9174}\textcolor{red}{8593}    &{\bf12.9174}\textcolor{red}{7} \\
$E_{exc}\left(4^{+}\right)$ &{\bf9.04142}\textcolor{red}{2878}    &{\bf9.04142}\textcolor{red}{5}\\
\hline
$B\left(E0:0^{+}\rightarrow0^{+}\right)$ &{\bf0.020567}\textcolor{red}{096}    &{\bf0.020567}\textcolor{red}{5}\\
$B\left(E1:0^{+}\rightarrow1^{-}\right)$ &{\bf12.80}\textcolor{red}{615358}    &{\bf12.80}\textcolor{red}{58} \\
$B\left(E2:0^{+}\rightarrow2^{+}\right)$ &{\bf0.3352}\textcolor{red}{04540}    &{\bf0.3352} \\
\hline \hline
\end{tabular}

\caption{Comparison of HFB and QRPA calculations performed for the nucleus
$^{18}$O without any pairing truncation and without Coulomb interaction.
For the Skyrme interaction we use the SLy4 parametrization \cite{SLy4}
with a delta (volume) pairing interaction \cite{HFODD} with strength
$\mathrm{V_{0}}=-200$ MeV(fm)$^{3}$ and a one-body center of mass
correction. The results are obtained with a harmonic oscillator basis
where the maximum oscillator shell included has principal quantum
number $N_{max}=5$ and the oscillator constant is set to $0.865$
(fm)$^{-1}$. In this table, the transition strengths are calculated
using the isoscalar transition operators of Ref.~\cite{T-operators}.
Energies have units of MeV and the $B\left(EI\right)$ transitions
are in units of $e^{2}\left(\mathrm{fm}\right)^{2I}$\label{tab:QRPA-Benchmark}.
For each multipolarity the state lowest in energy with an appreciable
strength is compared. When both codes give the same decimals, they
are printed in bold. }
\end{table}
Several different recipes on how to truncate the pairing space and
how to treat the Coulomb interaction exist in the literature, so in
order to make the benchmark results as useful as possible, the results
are obtained without any pairing truncation and without any Coulomb
interaction. The remaining parameters are listed in the caption of
Tab.~\ref{tab:QRPA-Benchmark}. 

The implementation based on the axially deformed \pr{HFBTHO} code
\cite{HFBTHO} allows us to test its accuracy by performing calculations
for excitations with different angular momentum projections on a principal
axis of the nucleus. Since the comparison is made for a spherical
nucleus, these different calculations should ideally give the same
result. In this way, for the QRPA implementation of \cite{HFBTHO},
the precision of the $2^{+}$ excitation was estimated to be roughly
$10^{-4}$ both for the energy and for the $B\left(E2\right)$ value.
As seen from Table \ref{tab:QRPA-Benchmark}, the lowest states calculated
with both codes agree to about this precision. A similar accuracy
test with the \textsf{\pr{HOSPHE}} code is not possible as it works
in spherical symmetry. But since \textsf{\pr{HOSPHE}} uses the same
mean-fields both in the QRPA and the HFB calculations we expect about
the same accuracy for the QRPA excitations as for the groundstate
energy. The full strength functions calculated with both codes were
also compared and turned out to be indistinguishable by eye when the
reduced transition probabilities are plotted as function of the energy
of the excited states.

\subsection{Iterative solutions}

As described above, the product of the QRPA matrix acting on an arbitrary
vector can be calculated without constructing the matrix explicitly.
When this technique is used, traditional matrix diagonalisation routines,
which need explicit information about the matrix elements can not
be used. Instead, one must resort to indirect iterative methods, such
as the Lanczos or Arnoldi \cite{arnoldi} methods. For non-hermitean
problems, such as the QRPA eigenvalue problem, the Implicitly Restarted
Arnoldi method (IRA) \cite{arnoldi-ira,arnoldi-saad} is one of the
most commonly used methods for finding accurate approximations for
the eigenstates lowest in energy. IRA is a more advanced version of
the original Arnoldi method, giving faster convergence and a reduction
in computational effort. As with the iterative Arnoldi method of \cite{TOI10},
the IRA method generates a set of basis vectors, usually called Ritz
vectors, which span a vector space called the Krylov subspace, and
uses these vectors to represent the QRPA eigenvectors. However, IRA's
use of restarting allows it to gradually improve the accuracy of a
set of eigenstates during iteration, using a reasonably small number
of Ritz vectors (typically a few hundred at most). 

In the extended \pr{HOSPHE} code, we use the numerical software \textsf{\pr{ARPACK}}
\cite{ARPACK}, which implements the IRA method. With this method,
the number of matrix-vector products needed in order to reach convergence
depends on the requested tolerance. Denoting the QRPA matrix $A$,
the approximate eigenvector $x$ and the corresponding approximate
eigenvalue $\lambda$, the iterations proceed until the accuracy measure
$\left\Vert Ax-\lambda x\right\Vert $ \cite{ARPACK} is less than
the requested tolerance. %
\begin{figure}[t]
\includegraphics[clip,width=0.95\columnwidth]{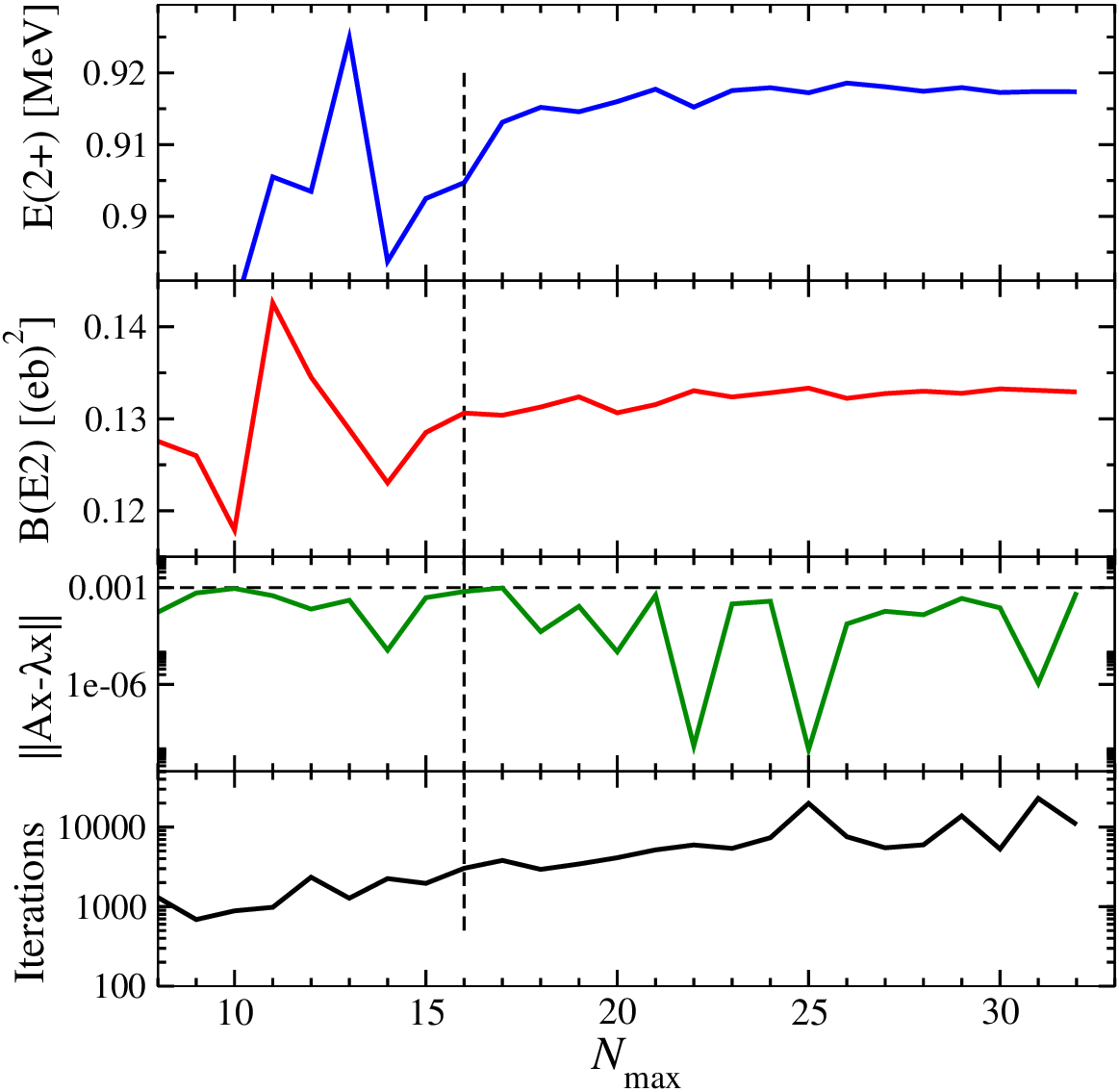}

\caption{\label{fig:Convergence}Convergence of the excitation energy and reduced
transition probability for the lowest $2^{+}$state in $^{214}$Pb
using the SLy4 interaction together with a separable Gaussian pairing
force \cite{seppair,GMR}. The convergence is shown as a function
of the maximum oscillator shell $N_{max}$ included in the basis.
The accuracy measure and the number of iterations needed to obtain
convergence are also shown. The tolerance parameter which determines
when to stop the Arnoldi iterations was set to 0.001. For $N_{max}=16$,
the wall clock time was 12 min on a desktop workstation (Intel Core
i7-2600K, 3.4GHz). In all cases the Krylov subspace was constructed
from 100 Ritz vectors which was estimated to give the fastest convergence
with $N_{max}=16$. }
\end{figure}

Fig.~\ref{fig:Convergence} shows the convergence as a function of
the basis size when applying the method for the calculation of the
lowest $2^{+}$ state in $\mathrm{^{214}Pb}$. The only truncation
employed is the number of main oscillator shells used for the basis.
As seen from this figure the accuracy measure is always lower than
the requested tolerance when the iterations finish and the number
of iterations required in order to reach the desired convergence increases
with the size of the basis. 

To find the lowest eigenstates a typical choice is to start from a
random initial guess (pivot) vector. For states with large transition
probabilities, the number of iterations needed can however be reduced
by instead starting from an initial pivot vector whose matrix elements
are set to the matrix elements of the corresponding electromagnetic
multipole operator \cite{TOI10}. In the case where pairing disappears
(and the numerical accuracy is high) the electromagnetic pivot also
filters out the states that have an overlap with the pivot and thus
removes states which correspond to pair addition or removal. Because
of these advantageous features we start from an electromagnetic pivot
in all the calculations presented.

For the calculations presented below, a value of 17 oscillator shells
($N_{max}=16$) was chosen to offer a good balance between accuracy
and computational speed. Assuming that the calculation with 33 oscillator
shells shown in Fig.~\ref{fig:Convergence} is fully converged, the
truncation error when stopping at 17 shells amounts to 0.01 $\mathrm{MeV}$
for the energy and 0.002 $\mathrm{\left(eb\right)^{2}}$ for the reduced
transition probability. Using 17 shells reduces the dimension of the
QRPA matrix to 8016 as compared to 59296 in the case of 33 shells.
With this smaller basis and using a tolerance parameter of 0.001,
the average time to calculate the lowest state for a nucleus in the
lead isotope chain is 6.5 min (Intel Core i7-2600K, 3.4GHz) and the
average number of iterations required is 2663.

Sets of a few lowest eigenvalues can also be extracted and requires
about the same number of iterations. For example, in in the case of
$^{192}$Pb, the number of iterations needed to extract 10, 20 and
30 positive energy eigenstates becomes 1974, 2648 and 3427 respectively.

\section{Numerical results}

\subsection{Influence of the pairing interaction}

In order to study the influence of the pairing interaction on $2^{+}$
states we compare the use of a zero-range delta interaction \cite{HFODD}
combined with a truncation in the equivalent spectra \cite{HFBTHO_CODE}
to the use of a separable Gaussian pairing force \cite{seppair,GMR}.
This force has a finite range and therefore does not need to be truncated.
In order to obtain reasonable pairing, the pairing strengths are tuned
to get the lowest quasiparticle energies to agree with the experimental
gaps extracted in Ref.~\cite{GAPvalues} using a four-point formula.
The resulting parameters obtained for the finite-range interaction
are shown in Table \ref{tab:pairing-constants}. %
\begin{table}[t]
\begin{tabular}{c c c} \hline \hline
Interaction&	$G_{n}$&	$G_{p}$\\ 
\hline 
SLy4&	655&	600\\ 
SKM*&	610&	550\\ 
SkX&	560&	530\\ 
\hline \hline
\end{tabular}

\caption{Strength parameters of the separable Gaussian pairing interaction
in units of MeVfm$^{3}$. For the range of the interaction we adopt
the value $a=0.660$ fm in all cases. \label{tab:pairing-constants} }
\end{table}

\begin{figure}[t]
\includegraphics[clip,width=0.85\columnwidth]{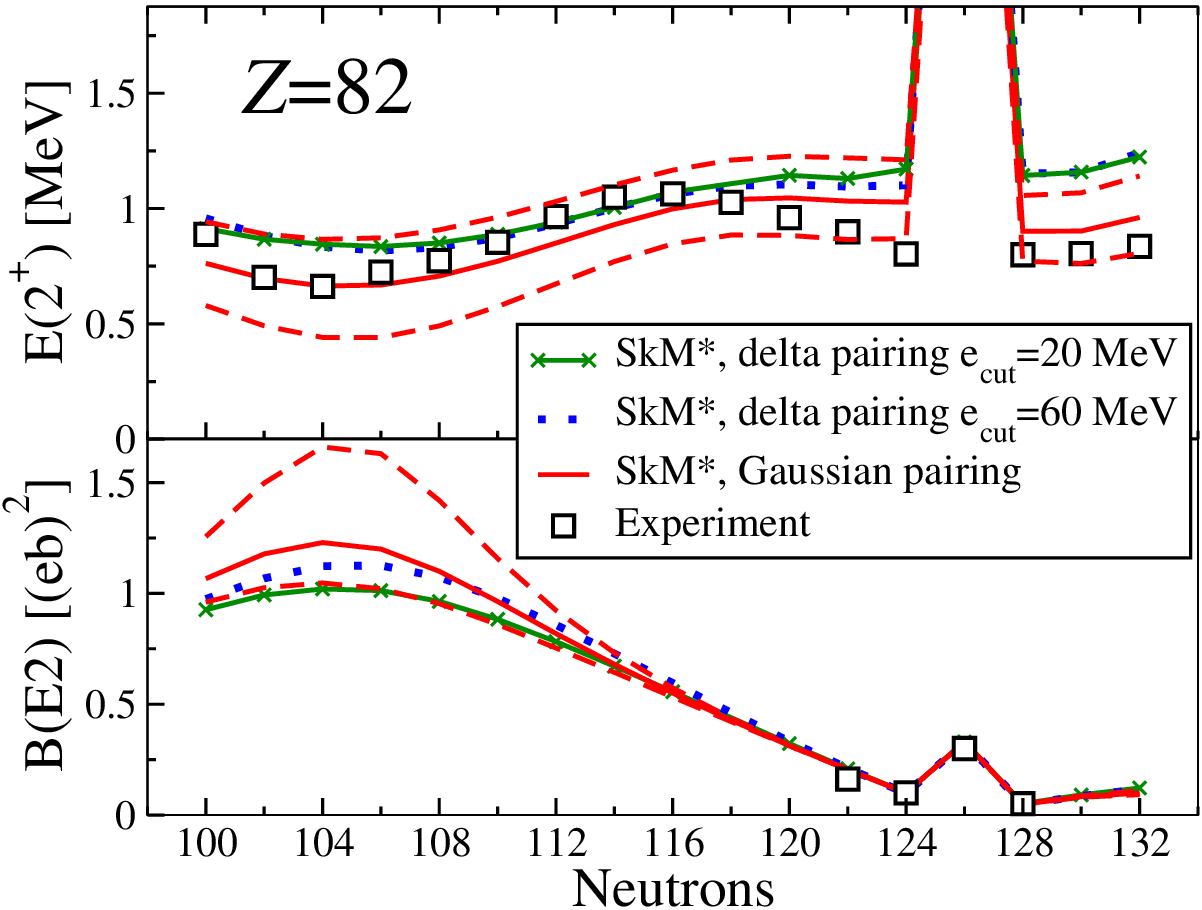}

\caption{\label{fig:pairing_test}Excitation energies and reduced transition
probabilities for Pb isotopes. Results are shown for different treatments
of the pairing interaction. The dashed lines denote the result of
changing the strength of the finite-range pairing with $\pm5\%$.
Decreasing the pairing lowers the energies and raises the $B\left(E2\right)$
values. The strength parameters for the zero-range interaction were
chosen as $V_{n}=-168$ and $V_{n}=-200$ (MeV(fm)$^{3}$) when the
cutoff in the equivalent spectra was taken as 60 MeV and 20 MeV respectively. }
\end{figure}

Results for the lead isotopes using the different pairing interactions
are shown in Fig.~\ref{fig:pairing_test}. %
As seen in this figure there are fluctuations in the energies which
depend on the choice of pairing force. Comparing the two pairing interactions,
it appears that the finite-range interaction is slightly better in
capturing the fluctuations of the experimental energies.

In the equivalent spectra method the normal and abnormal density-matrices
are truncated during the HFB iterations \cite{HFBTHO_CODE}. However,
in the subsequent QRPA calculation we used non-truncated wavefunctions
without any energy cut for the residual particle-particle interaction.
This way of using the equivalent spectra method is therefore slightly
inconsistent and a better truncation recipe is desired. In the following
we will only use the finite-range pairing interaction which does not
need to be truncated and allows us to treat HFB and QRPA in a consistent
way. 

The effect of changing the strength of the Gaussian pairing interaction
with $\pm5\%$ is shown in Fig.~\ref{fig:pairing_test} with dashed
lines. Both the energies and the transitions are sensitive to such
a change, i.e. a decreese of the pairing lowers the excitation energies
and raises the $B\left(E2\right)$ values. The effect is seen to be
largest for $N=104$ which is just between two magic numbers. The
schematic fits of the pairing strengths appear to give quite reasonable
values for the lead isotopes with an average $E_{2^{+}}$ energy that
agrees roughly with experiment.

\subsection{$J^{\pi}=2^{+}$ states in Pb and Sn isotopes}

In this work we consider three different Skyrme parameterizations:
SkM{*}, SLy4 and SkX. SKM{*} is based on the SkM parameters \cite{skm},
but has been adjusted further using results from fission barrier calculations
\cite{skm*}. The original SkM parameters were determined by considering
both static ground state properties as well as some dynamical properties
including monopole and quadrupole resonances \cite{skm}. SLy4 was
adjusted with special care taken to model neutron matter in order
to facilitate the description of neutron rich nuclei \cite{SLy4}.
The accuracy of QRPA based on these interactions was recently studied
and compared to calculations based on the generator-coordinate method
GCM \cite{2+states-JT}. It was found that SkM{*} reproduced experimental
$2^{+}$ states more accurately than SLy4 and the QRPA results were
similar to results obtained with the GCM. In addition, we also consider
the SkX interaction which has been tuned with special focus on reproducing
single-particle states in double-magic nuclei \cite{skx}. 

\begin{figure}[t]
\includegraphics[clip,width=1\columnwidth]{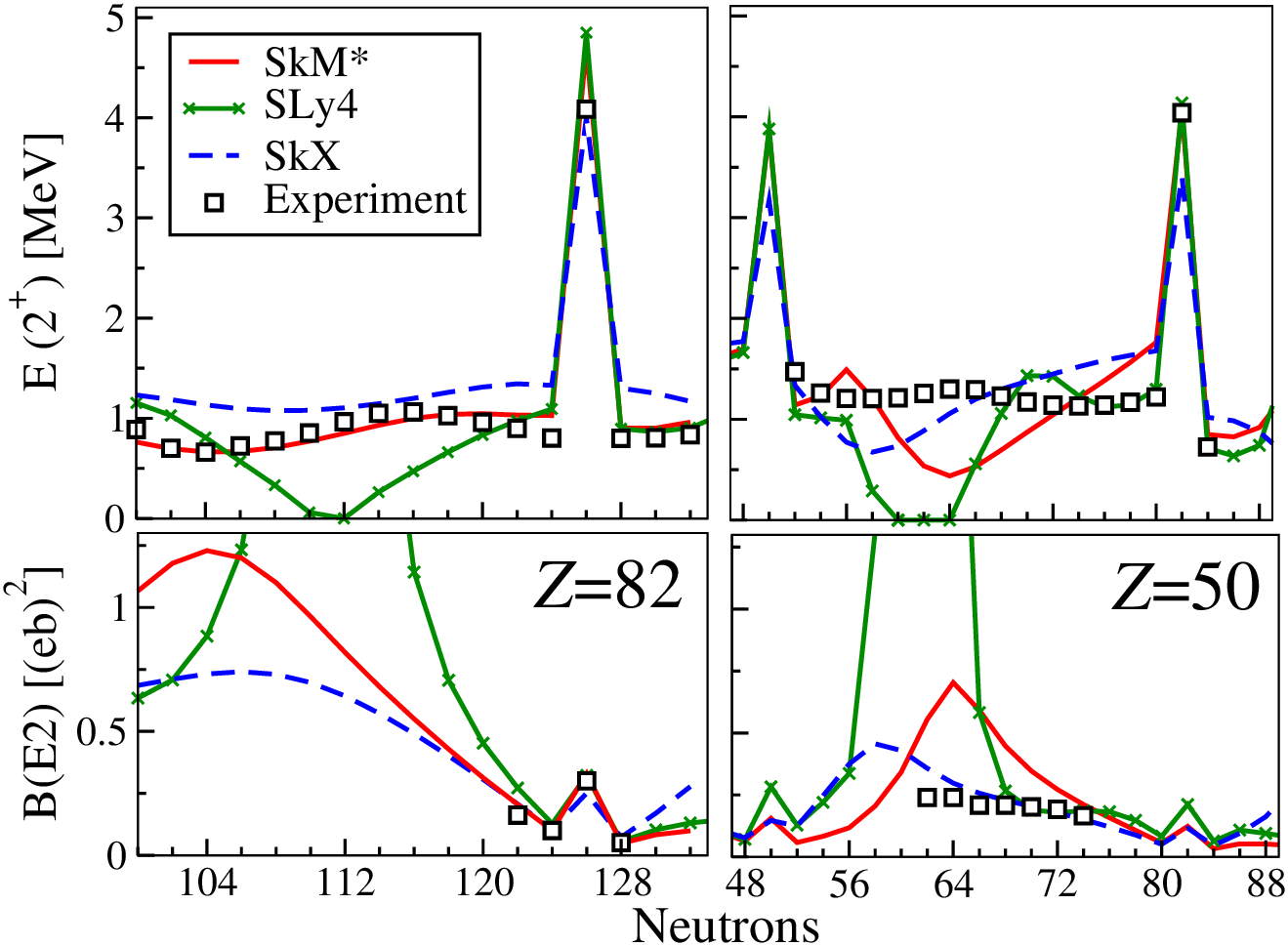}

\caption{\label{fig:skyrm}Excitation energies and reduced transition probabilities
($B(E2;0_{g.s.}^{+}\rightarrow2_{1}^{+})$) for Pb and Sn isotopes.
Results are shown for three different Skyrme parameterizations. The
experimental values are taken from Ref. \cite{data2p}. }
\end{figure}

Results for the lead isotopes using the three different Skyrme interactions
are shown in the left hand panels of Fig.~\ref{fig:skyrm}. %
For the double magic nucleus Pb$_{126}$, SkX gives the correct $2^{+}$
energy while the other two forces overestimate this excitation energy.
As the neutron number is reduced, the predictions show considerable
differences. SLy4 gives zero energy solutions around $N=112$, indicating
a transition to a deformed ground state, while the other two interactions
appear to be stiffer towards deformation and give more realistic results. 

\begin{figure}[t]
\includegraphics[clip,width=0.67\columnwidth]{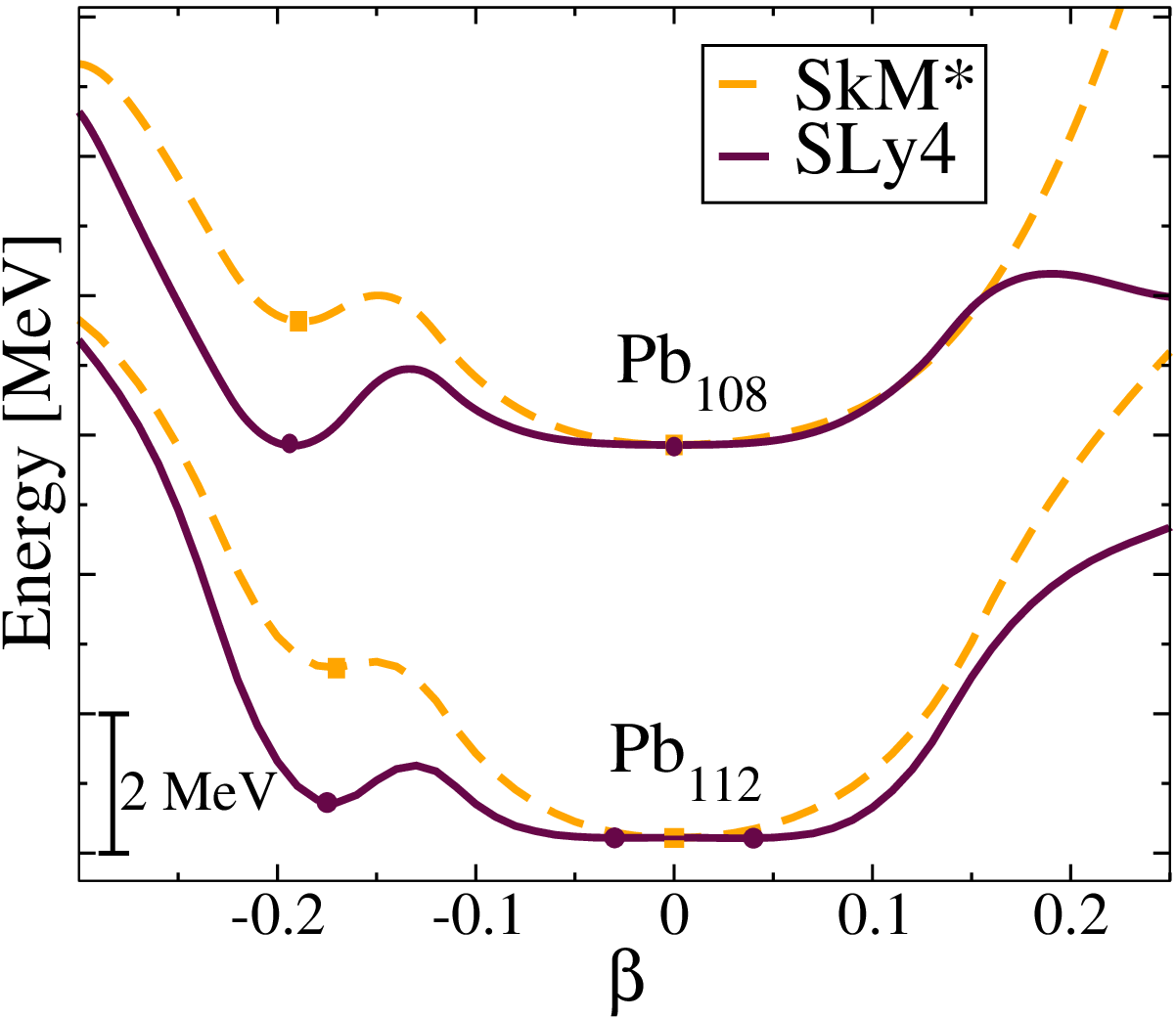}

\caption{\label{fig:pb_surf}HFB ground-state energy as a function of quadrupole
deformation $\beta$ \cite{HFBTHO_CODE} for Pb$_{108}$ and Pb$_{112}$.
The calculation was performed using the \pr{HFBTHO} code \cite{HFBTHO_CODE}.
The local minimas are marked with symbols. Constant shifts of both
curves have been applied in order to make them fit in the figure.}
\end{figure}

The ground-state energies of Pb$_{108}$ and Pb$_{112}$ as a function
of quadrupole deformation are shown in Fig.~\ref{fig:pb_surf}. %
The energy curves are calculated using delta pairing instead of the
Gaussian pairing which means that the curves are slightly inconsistent
with the QRPA calculations, but the general features will be the same.
As seen from this figure, both interactions predict spherical minima
for Pb$_{108}$, although the lowest minimum with SLy4 is the oblate
one with quadrupole deformation $\beta=-0.19$. The spherical minimum
obtained with SkM{*} is stiffer than with SLy4 which is probably the
reason why the $2^{+}$ energy is predicted higher. As one moves to
Pb$_{112}$, SLy4 gives the spherical point as a maximum with neighboring
slightly deformed minima. In this case our QRPA calculation is likely
to give a zero energy solution as the assumed spherical ground state
is no longer stable with respect to quadrupole deformations. The $2^{+}$
energies obtained with the SkM{*} interaction agree rather well with
experiment and seem to favor the prediction of stiffer energy surfaces.

In the QRPA formalism, an expression for the operator which creates
the excited states by acting on the QRPA ground state can be written
as \[
O_{\alpha}^{\dagger}=\sum_{k<k'}Z_{kk'}^{\alpha}\alpha_{k}^{\dagger}\alpha_{k'}^{\dagger}-Z'{}_{kk'}^{\alpha*}\alpha_{k'}\alpha_{k}.\]
 In order to discuss the structure of the solutions, we label the
$kk'$ components of this creation operator using the quantum numbers
of the quasiparticle operators. As an example, if both $k$ and $k'$
refer to a quasiproton (quasineutron) in a $i_{13/2}$ shell, the
corresponding component is denoted as $\pi(i_{13/2})^{2}$ ($\nu(i_{13/2})^{2}$).
Indeed, in the limit when $Z'{}_{kk'}^{\alpha}=0$ this turns into
the usual notation for writing two quasiprotons (quasineutrons) in
the $i_{13/2}$ shell. 

The major oscillator $N_{osc.}$ quantum number is not preserved in
the calculations, but for simplicity we will refer to the mixed orbitals
using the harmonic oscillator ordering. For example the lowest $p_{3/2}$
and $p_{1/2}$ quasiparticle orbitals will be referred to as being
of $N_{osc}=1$ character, although these orbitals also contain contributions
from higher oscillator shells. 

\begin{figure}[t]
\includegraphics[clip,width=0.95\columnwidth]{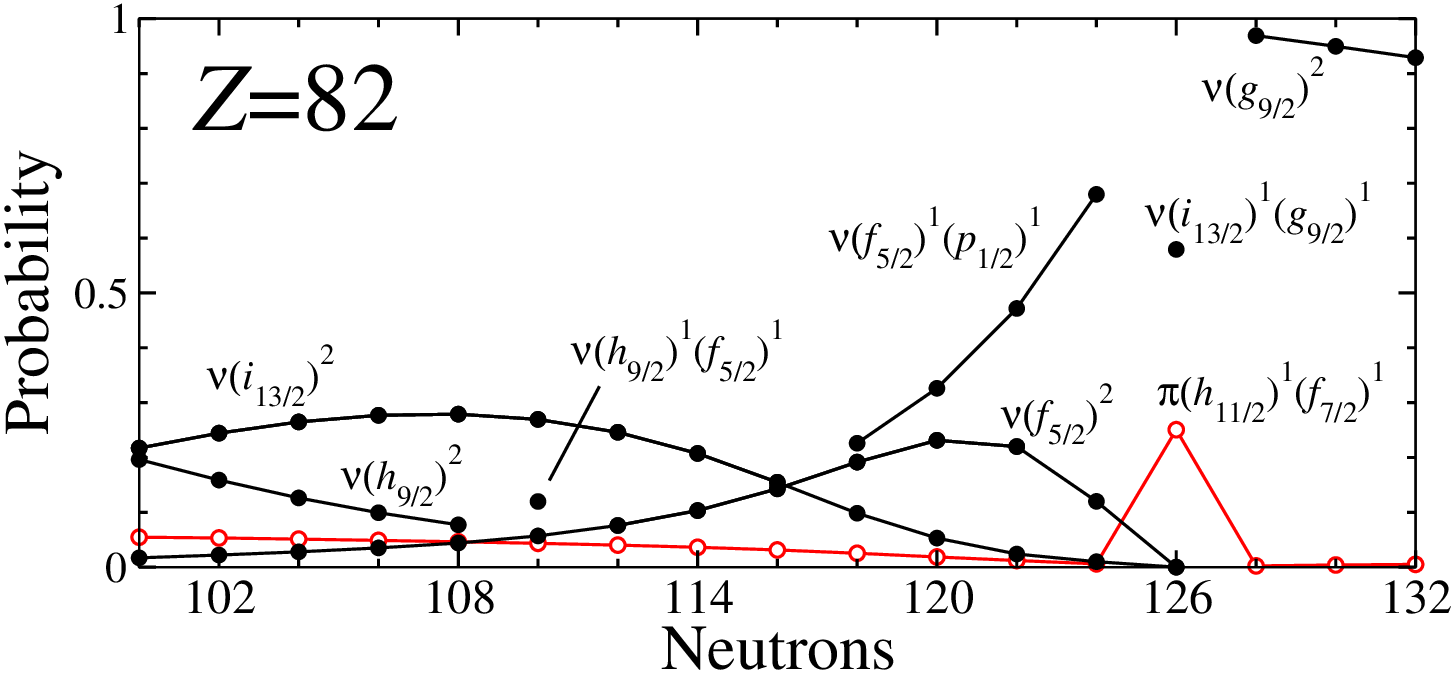}

\caption{\label{fig:pb_struc}Structure of QRPA $2_{1}^{+}$ states in lead
isotopes using SkM{*}. }
\end{figure}

\begin{figure}[t]
\includegraphics[clip,width=0.9\columnwidth]{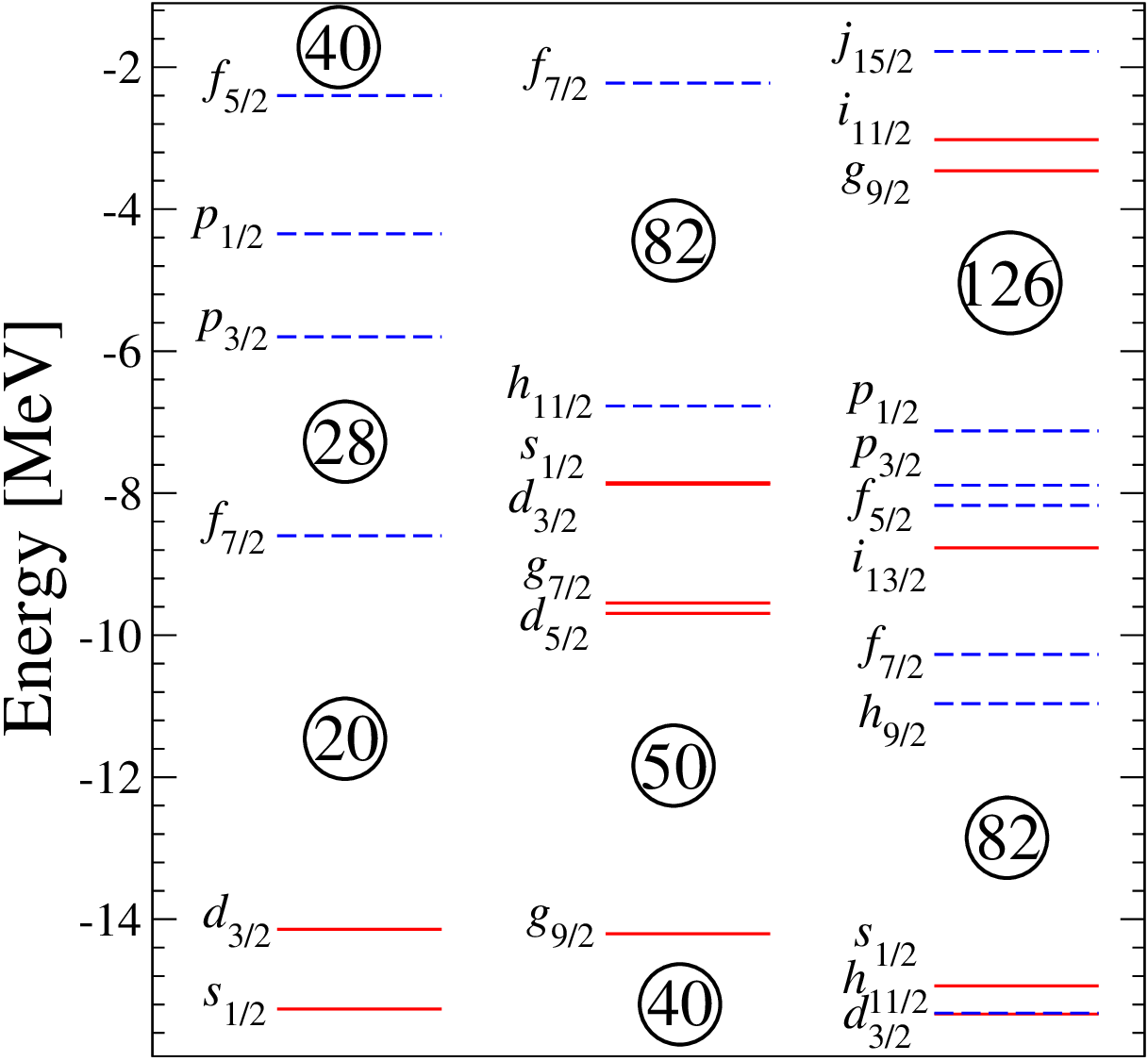}

\caption{\label{fig:sp-levs}Theoretical (SkX) neutron single-particle levels.
The left set of levels is for Ca$_{28}$, the middle set for Sn$_{82}$,
and the right set is for Pb$_{126}$. Positive parity levels are shown
with full lines and negative parity levels with dashed lines. }
\end{figure}

We denote the probability $P_{\alpha j,\alpha'j'}$ for different
components in the wavefunctions of the excited states by summing contributions
from the different $m$ quantum numbers as \[
P_{\alpha j,\alpha'j'}=\sum_{m,m'}|Z_{\alpha jm,\alpha'j'm'}|^{2}-|Z'_{\alpha jm,\alpha'j'm'}|^{2}.\]
 Defined in this way, the largest components in the calculated $2_{1}^{+}$
states in the chain of lead isotopes are shown in Fig.~\ref{fig:pb_struc}.
As seen in this figure, the largest proton and neutron components
in Pb$_{126}$ involve particle-hole excitations across the $Z=82$
and $N=126$ gaps. In the other isotopes, where neutron pairing is
active, the $2^{+}$ states mainly involve neutron excitations ($\sim85\%$)
with the dominating component being 15-30 \% $\nu\left(i_{13/2}\right)^{2}$
for $N=100-116$. The calculated neutron single-particle levels are
shown in Figure \ref{fig:sp-levs}%
 and as expected the dominating components in the $2^{+}$ states
involve excitations among the shells close to the Fermi-level. 

It should be noted that the transition strenghts are calculated directly
from the electromagnetic operators \cite{Ring80} without any effective
charges. Therefore it is the smaller proton components, suppressed
because of the magic proton number, which determine the electromagnetic
properties. It is also interesting to notice that above the $126$
gap, the $2_{1}^{+}$ states are composed of rather pure two-quasi-neutron
excitations to the $g_{9/2}$ shell. 

\begin{figure}[t]
\includegraphics[clip,width=0.8\columnwidth]{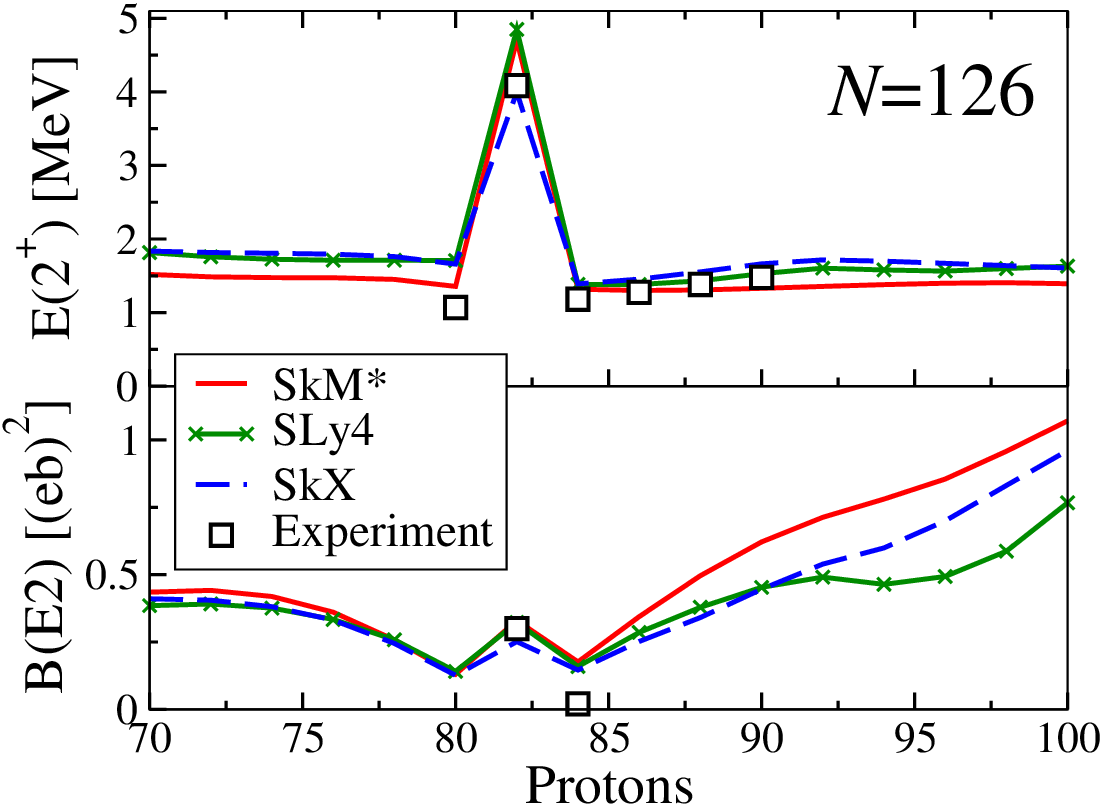}

\caption{\label{fig:pb_isotones}Same as in Fig.~\ref{fig:skyrm}, but for
$N=126$ isotones. }
\end{figure}

For the $N$=126 isotones shown in Fig.~\ref{fig:pb_isotones} there
is a similar accuracy as obtained for the lead isotopes. %
To have a better idea about the amount of collectivity that should
be present when going away from $_{82}$Pb$_{126}$, more experimental
transition probabilities are clearly needed and experiments to measure
the unknown $B\left(E2\right)$ values for the isotopes $_{82}\mathrm{Pb_{114,116,118,120}}$,
$_{78}$Pt$_{122,124}$ and $_{80}$Hg$_{126}$ are planned \cite{experiments}. 

Results for the Sn chain are shown in the right hand panels of Fig.~\ref{fig:skyrm}.
As in the case of the Pb chain, the SLy4 interaction gives some zero
energy solutions, while the other two interactions produce dips in
the excitation energies for some isotopes, but do not reach zero.
For the two double magic Sn isotopes, the excited states are calculated
to be roughly even mixture of proton and neutron excitations, while
the excited states in the semi-magic isotopes mainly involve neutron
excitations. With SkX, the largest components are $\nu\left(d_{5/2}\right)^{2}$
for Sn$_{52-54}$, $\nu\left(g_{7/2}\right)^{2}$ for Sn$_{56-64}$
and $\nu\left(h_{11/2}\right)^{2}$ for Sn$_{66-80}$. The positions
of these shells (shown in Figure \ref{fig:sp-levs}) are thus important
in order to reproduce the details of the experimental data.

\subsection{$J^{\pi}=2^{+}$ states in Ni and Ca isotopes}

\begin{figure}[t]
\includegraphics[clip,width=1\columnwidth]{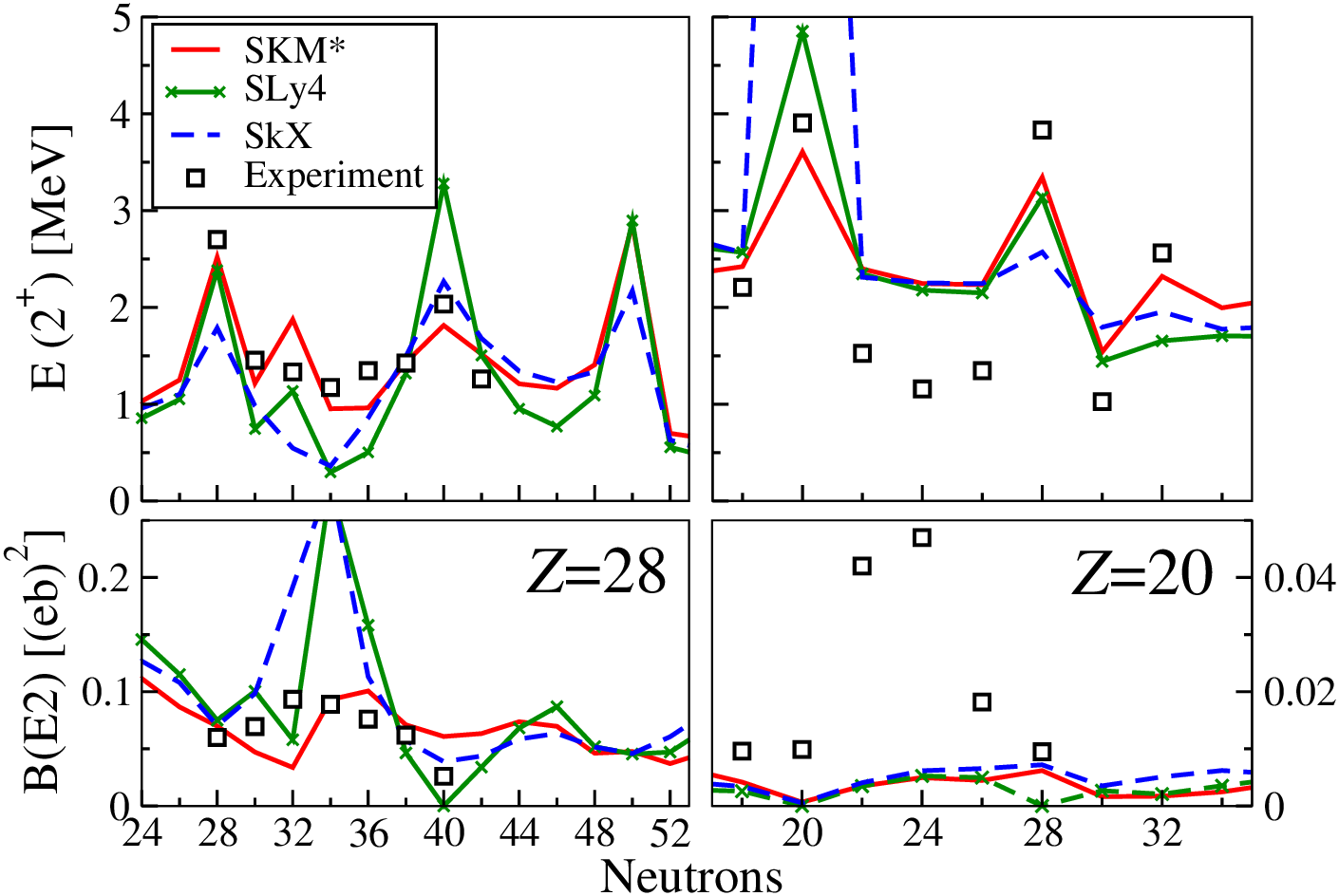}

\caption{\label{fig:skyrm-1}Same as in Fig.~\ref{fig:skyrm}, but for Ni
and Ca isotopes. }
\end{figure}

The results for the Ni chain are shown in the left hand panels of
Fig.~\ref{fig:skyrm-1}. %
In Ni$_{28}$, the $2^{+}$ state is built from an almost equal mixture
of proton and neutron excitations , while the $2^{+}$ states in the
other nickel isotopes are dominated by neutron excitations. SkX predicts
a smaller 28 gap than the other interactions and a $2^{+}$ state
in Ni$_{28}$ which is lower than in experiment. SkX also predicts
a smaller gap at $N$=32 and does not show the spike in excitation
energy obtained with the other interactions for Ni$_{32}$. 

QRPA calculations based on the relativistic-mean-field model for Ni$_{40}$
\cite{RMF-low-lying} overpredicted the energy of the $2^{+}$ state
by roughly three times the experimental energy. A suggested explanation
was missing 2p-2h and higher order excitations among the neutrons
\cite{RMF-low-lying}. Since the neutron Fermi-level is located between
opposite parity shells, this state will be overpredicted whenever
the neutron pairing goes to zero. With SLy4 we also obtain an overprediction,
although less severe, while the other interactions predict excitation
energies close to the experimental value. It is interesting to note
that with SkX the ground state is calculated to have an average gap
\cite{GAPS} of $\Delta_{n}$ = 1.38 MeV and the $2^{+}$ state is
obtained with both correct energy and transition strength. The state
is built as a mixture of proton ($\sim23\%$) and neutron ($\sim77\%$)
excitations, where the largest component is $\nu\left(g_{9/2}\right)^{2}$. 

Results for the Ca chain are shown in the right hand panels of Fig.~\ref{fig:skyrm-1}.
For Ca$_{20}$ one should notice that the Fermi-levels for both neutrons
and protons are right between shells with opposite parity. Therefore,
if pairing disappears, as happens with SkX, the lowest particle-hole
excitations with positive parity are between shells of $N_{osc}=2$
and $N_{osc}=4$ character, and the excitations have to bridge an
energy gap of around 15 MeV, which pushes the predicted $2^{+}$ state
up to a high energy. With SkM{*} (SLy4) there is some pairing remaining
and a low-lying $2^{+}$ state can be constructed with a dominating
$\pi\left(d_{3/2}\right)^{2}$ ($\pi\left(f_{7/2}\right)^{2}$) quasiparticle
component. This state is likely to have an average particle number
which is somewhat wrong, but in this work we only remove the excitations
being in the wrong nucleus if pairing vanishes completely. Notice
also that since the transition operator is of particle-hole type,
excitations corresponding to addition or removal of two particles
gives zero for the transition strength. 

In general, the $2^{+}$ states in the Ca chain are predicted to have
too little collectivity compared to experiments. The states in Ca$_{22-26}$
are predicted to be rather pure excitations within the $f_{7/2}$
shell. For example, with the SkX interaction the component of $\nu\left(f_{7/2}\right)^{2}$
is 95, 90, and 82\% for the excitations in Ca$_{22-26}$ respectively.
In order to induce more collectivity, it appears likely that proton
two-particle-two-hole excitations across the $Z=20$ gap must be explicitly
included which goes beyond the present QRPA treatment.

\subsection{$J^{\pi}=3^{-}$ states in Pb and Sn isotopes}

\begin{figure}[t]
\includegraphics[clip,width=1\columnwidth]{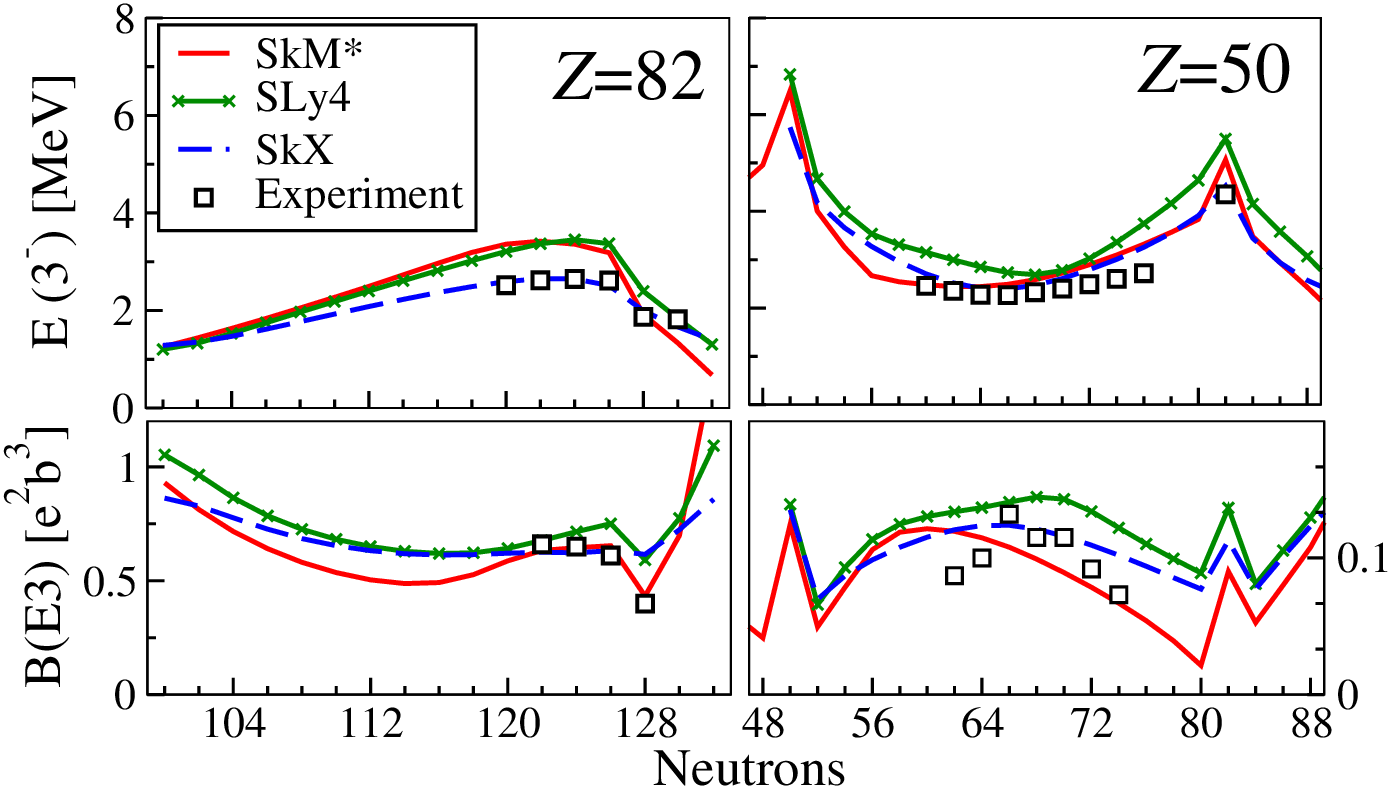}

\caption{\label{fig:3m_high}Excitation energies and reduced transition probabilities
($B(E3;0_{g.s.}^{+}\rightarrow3_{1}^{-})$) for the lowest $3^{-}$
states in Pb and Sn isotopes. Results are shown for three different
Skyrme parameterizations. The experimental data for the $3^{-}$ states
are taken from the compilation \cite{data3m}. }
\end{figure}

The calculated $3^{-}$ states for lead isotopes are shown in the
left hand panels of Fig.~\ref{fig:3m_high}. %
The most striking feature of the data is the dip in excitation energy
seen when going from Pb$_{126}$ to Pb$_{128}$. In Pb$_{126}$ the
$3^{-}$ state is created by an roughly equal amount of neutron and
proton excitations across the 82 and 126 gaps. When two more neutrons
are added, negative parity states can be made by exciting particles
between the neighboring positive parity $g_{9/2}$ shell and negative
parity $j_{15/2}$ shell, located above the 126 gap (see Figure \ref{fig:sp-levs}).
Therefore the $3^{-}$ state in Pb$_{128}$ can be built at a low
cost mainly from neutron excitations, which explains the dip in the
experimental energy. As seen in Fig.~\ref{fig:3m_high}, SkX reproduces
the experimental lowest $3^{-}$ energies and $B(E3)$ values almost
perfectly except for the dip in the transition probability seen for
Pb$_{128}$. The forces SkM{*} and SLy4 have a less perfect overall
agreement, but SkM{*} agrees with experiment in the case of $\mathrm{Pb_{128}}$.

In $\mathrm{Pb_{100}}$ with SkX, the excitation operator for the
lowest $3^-$ state consists of 17 \% proton excitations, where the
largest component is only 6 \% $\pi(d_{3/2})^{1}(h_{9/2})^{1}$ and
the main neutron component is 62 \% $\nu(f_{7/2})^{1}(i_{13/2})^{1}$.
The $3^{-}$ states of the neutron deficient Pb isotopes have many
small proton components contributing. When the neutron number increases,
proton excitations become more dominant, but with fewer components
contributing. At \emph{N}=124, proton excitations constitute 48 \%
with one dominant component, 29 \% $\pi(d_{3/2})^{1}(h_{9/2})^{1}$,
while the largest neutron component is 13 \% $\nu(p_{3/2})^{1}(g_{9/2})^{1}$.
Thus, with QRPA, we see a transition from strong proton configuration
mixing in $\mathrm{Pb}_{100}$ to strong neutron configuration mixing
in $\mathrm{Pb}_{124}$. 

The lowest experimental and calculated $3^{-}$ states of tin isotopes
are shown in the right hand panels of Fig.~\ref{fig:3m_high}. The
energies are reproduced almost perfectly in the region \emph{N}=60-72,
using SkX and SkM{*} forces. Closer to the\emph{ N}=82 shell closure,
the predicted energies become too high. The experimental $B(E3)$
values are reproduced roughly by all three Skyrme forces, but none
of the three forces gives a truly accurate description of the finer
details. The excitations are dominated by the $\nu(d_{5/2})^{1}(h_{11/2})^{1}$
configuration which has a component of 70 \% in $\mathrm{Sn}_{52}$,
and gradually goes down to 50 \% in $\mathrm{Sn_{80}}$. The second
largest excitation is 3\% $\nu(g_{9/2})^{1}(h_{11/2})^{1}$ in $\mathrm{Sn}_{52}$,
but starting from $\mathrm{Sn}_{54}$ the second largest excitation
is $\nu(g_{7/2})^{1}(h_{11/2})^{1}$ and its amplitude grows steadily
from 5\% to 10\% in $\mathrm{Sn}_{80}$. Other neutron excitations
between the $\nu h_{11/2}$ subshell and the $N_{osc}=4$ orbitals
are excluded because of angular momentum selection rules. The proton
fraction of the transitions is an almost constant 10 \% and is composed
of many small-amplitude excitations. The lowest energies and the largest
transitions strengths are obtained for $N$=66 when the neutron Fermi-level
is close to the $\nu h_{11/2}$ intruder shell and negative parity
excitations correspond to a low energy cost.

\subsection{$J^{\pi}=3^{-}$ states in Ni and Ca isotopes}

\begin{figure}[t]
\includegraphics[clip,width=1\columnwidth]{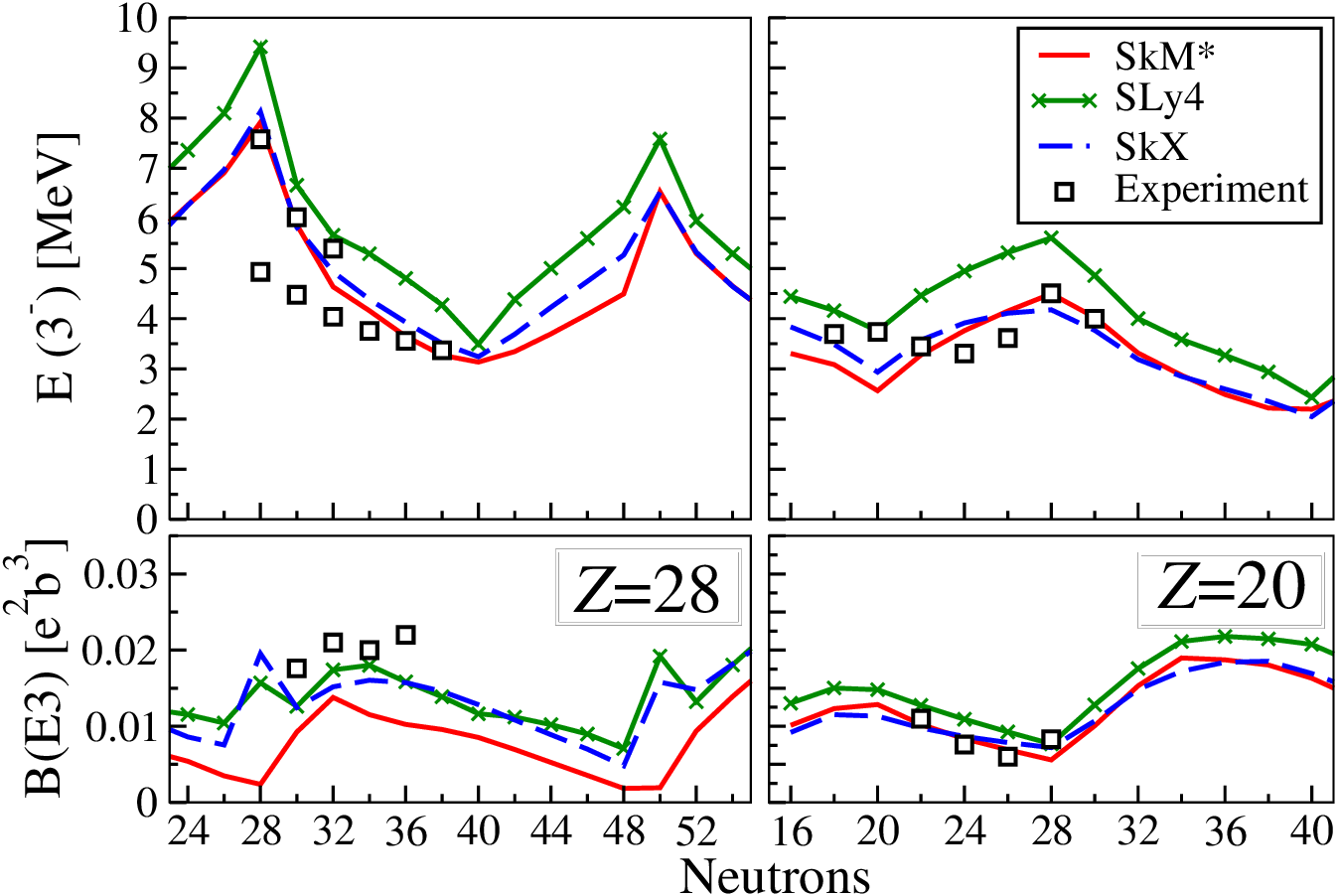}

\caption{\label{fig:3m_zsmall}Same as in Fig.~\ref{fig:3m_high}, but for
Ni and Ca isotopes. }
\end{figure}

For the Ni isotopes shown in the left hand panels of Fig.~\ref{fig:3m_zsmall}
we also show experimental data for the second $3^{-}$ states in Ni$_{28-32}$
taken from \cite{ensdf}. %
For some reason the calculations for $\mathrm{Ni}_{28-32}$ agree
better with the second $3^{-}$ states. Especially the lowest $3^{-}$
state in $\mathrm{N}\mathrm{i}_{28}$ is predicted about 3 MeV too
high in energy. It should be noted that this state has not been clearly
identified as $3^{-}$ in experiments \cite{data3m}, contrary to
the $3^{-}$ states in the other Ni isotopes. One should also note
that with Skyrme type interactions one usually neglects proton-neutron
pairing and part of the isovector particle-hole interaction which
may have an influence around the \emph{N=Z} line. With SkX, the lowest
theoretical $3^{-}$ state in $\mathrm{Ni}_{28}$ is calculated to
be of isoscalar type. This state is built from 28 \% $\nu(d_{3/2})^{1}(p_{3/2})^{1}$
and 22 \% $\pi(d_{3/2})^{1}(p_{3/2})^{1}$ along with smaller probability
excitations to the orbitals of $N_{osc}$= 4 character.

For the semi-magic nickel isotopes with \emph{N}=30-48, the $3^{-}$
states are dominated by neutron excitations, $\nu(p_{3/2})^{1}(g_{9/2})^{1}$
which decrease from 75 \% in \emph{N}=30 to 16 \% in \emph{N}=48
and $\nu(f_{5/2})^{1}(g_{9/2})^{1}$, which increase from 2 \% in
\emph{N}=30 to 74 \% in \emph{N}=48. The proton components are small
since it is more favorable to excite neutrons when the proton Fermi-level
is just above $f_{7/2}$. The double magic Ni$_{50}$ has a different
structure than the semi-magic isotopes, the dominating component being
36 \% $\nu(p_{1/2})^{1}(d_{5/2})^{1}$.

The results for calcium isotopes are shown in the right hand panels
of Fig.~\ref{fig:3m_zsmall}. In this case, the lowest experimental
$3^{-}$ energies are best reproduced by SkM{*} and SkX while SLy4
gives too high energies. However, in general the finer details of
the $3_{1}^{-}$ energies between \emph{N}=20 and \emph{N}=26 are
not reproduced. Especially for the $N=Z$ nucleus $\mathrm{Ca}_{20}$,
the energy calculated with SkM{*} and SkX becomes too low. 

\begin{figure}[t]
\includegraphics[clip,width=0.95\columnwidth]{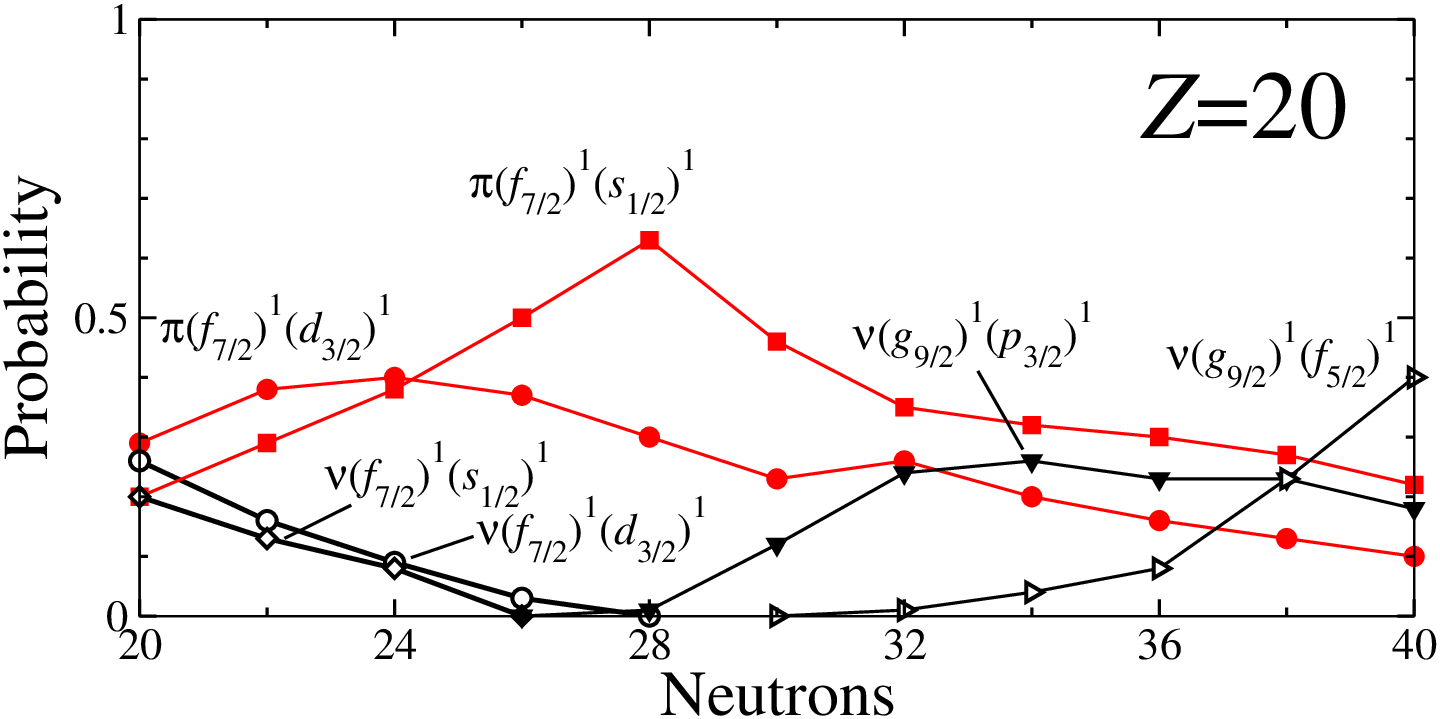} 

\caption{\label{fig:3m_zsmall-1}Leading two-quasiparticle components of QRPA
$3_{1}^{-}$ states in Ca isotopes using SkX. }
\end{figure}

The leading wave function components of the calcium isotopes are shown
in Fig.~\ref{fig:3m_zsmall-1}. %
As seen in this figure, the $3^{-}$ state in $_{20}\mathrm{Ca}_{20}$
is composed of a fairly even mixture of proton and neutron excitations
from the $d_{3/2}$ and $s_{1/2}$ shells below the 20 gap to the
$f_{7/2}$ shell just above the gap (see Figure \ref{fig:sp-levs}).
Going towards $\mathrm{Ca}_{28}$ the neutron excitations become suppressed
as the Fermi-level reaches the middle of the $N_{osc}=3$ shell and
proton excitations start to dominate. When more neutrons are added,
neutron excitations to the $g_{9/2}$ shell start to appear and become
the largest components in Ca$_{40}$.

\section{Summary and conclusions}

An iterative method for the solution of the QRPA equations which avoids
the construction of the large QRPA matrix was employed for the calculation
of low-lying vibrational states. The method uses the Implicitly Restarted
Arnoldi approach for the solution of the non-hermitian eigenvalue
problem. In this approach, only the action of the matrix on a Ritz
vector is needed. As demonstrated, this can be expressed in terms
of the fields generated by the transitional densities corresponding
to the Ritz vector. Our study shows that the method is numerically
stable and typically requires a few thousand iterations in order to
produce well converged lowest eigenstates. 

The new solution method was applied to the calculation of excitation
energies and decay rates of the first $2^{+}$ and $3^{-}$ vibrational
states in a set of spherical even-even nuclei. The calculations were
performed using three different Skyrme interactions together with
a finite-range pairing force. Overall a quite reasonable agreement
with experimental data was obtained. The main difficulties seem to
be in the description of $2^{+}$ states right between two magic neutron
numbers where the different interactions tend to give different results
and even zero energy solutions. 

Difficulties were also observed for the Ca isotopes where all the
interactions gave too little collectivity. These difficulties are
probably related to the limitations of the QRPA method itself, as
it only includes two-quasiparticle excitations. However, since our
method is rather computationally inexpensive, it may become practical
to consider extensions of QRPA, for example higher-order QRPA approaches
or boson expansion methods, which allow to treat more complicated
excitations that could improve the results. 

Because of the low numerical cost and low memory requirements, the
method appears promising for applications to deformed nuclei where
the dimensions become substantially larger. Indeed, the methods of
this work are quite analogous to the ones used in the nuclear Shell
Model community, where quite similar iterative methods have been used
for large-dimensional hermitean eigenvalue problems. However, contrary
to the Shell Model where the dimensions increase exponentially and
multi-major shell calculations for heavy nuclei are almost impossible,
the QRPA method stays tractable. Therefore, as a next step, we will
implement the method in a code able to treat nuclei with deformed
ground states.

For spherical nuclei, the speed of the iterative method opens the
possibility to directly compute low-lying states and include them
as part of the observables used when fitting the parameters of new
improved Skyrme interactions. However, care must be taken to analyse
the structure of the included states in order to ensure that a QRPA
description is compatible with the experimental states.

\section*{acknowledgments}

B.G. Carlsson acknowledges D. DiJulio for discussions concerning the
experimental data and the Royal Physiographic Society in Lund for
providing funding for the computers on which the calculations were
performed. We also thank I. Ragnarsson for valuable comments on the
manuscript. This work was supported in part by the Academy of Finland
and University of Jyväskylä within the FIDIPRO programme.

\end{document}